\documentclass{jaa}
\usepackage[numbers, sort, comma, square]{natbib}
\usepackage{graphicx}

\begin{document}\sloppy

\title{Multi-wavelength view of the galactic black-hole binary GRS 1716--249}


\author{Sandeep K. Rout \textsuperscript{1,2,*}, Santosh V. Vadawale\textsuperscript{1}, Aarthy E.\textsuperscript{1,2}, Shashikiran Ganesh\textsuperscript{1}, Vishal Joshi\textsuperscript{1}, Jayashree Roy\textsuperscript{3}, Ranjeev Misra\textsuperscript{3} \and J. S. Yadav\textsuperscript{4}}
\affilOne{\textsuperscript{1}Physical Research Laboratory, Navarangpura, Ahmedabad 380009, India.\\}
\affilTwo{\textsuperscript{2}Indian Institute of Technology, Palaj 382355, India.\\}
\affilThree{\textsuperscript{3}Inter-University Center for Astronomy and Astrophysics, Ganeshkhind, Pune 411007, India\\}
\affilFour{\textsuperscript{4}Department of Physics, Indian Institute of Technology, Kanpur 208016, India.}


\twocolumn[{

\maketitle

\corres{skrout@prl.res.in}


\begin{abstract}

The origins of X-ray and radio emissions during an X-ray binary outburst are comparatively better understood than those of ultraviolet, optical and infrared radiation. This is because multiple competing mechanisms - emission from intrinsic \& irradiated disk, secondary star emission, Synchrotron emission from jet and/or non-thermal electron cloud, etc - peak in these mid-energy ranges. Ascertaining the true emission mechanism and segregating the contribution of different mechanisms, if present, is important for correct understanding of the energetics of the system and hence its geometry and other properties. We have studied the multi-wavelength spectral energy distribution of the galactic X-ray binary GRS 1716--249 ranging from near infrared ($5 \times 10^{-4}\, keV$) to hard X-rays ($120\, keV$) using observations from \emph{AstroSat}, \emph{Swift}, and Mount Abu Infrared Observatory. Broadband spectral fitting suggests that the irradiated accretion disk dominates emission in ultraviolet and optical regimes. The near infrared emission exhibits some excess than the prediction of the irradiated disk model, which is most likely due to Synchrotron emission from jets as suggested by radio emission. Irradiation of the inner disk by the hard X-ray emission from the Corona also plays a significant role in accounting for the soft X-ray emission.

\end{abstract}

\keywords{Black holes---X-ray binary---NIR/optical/UV---Synchrotron emission---Jets}

}]


\doinum{12.3456/s78910-011-012-3}
\artcitid{\#\#\#\#}
\volnum{000}
\year{0000}
\pgrange{1--}
\setcounter{page}{1}
\lp{1}

\section{Introduction} \label{introd}

During outburst, a low mass X-ray binary brightens by several orders of magnitude in the entire electromagnetic spectrum. While the origins of radio and X-ray emission are extensively studied, the mid-energy emission in ultra-violet (UV), optical, and infrared (IR) poses a certain level of ambiguity (van Paradijs \& McClintock 1995; Charles \& Coe 2006). Soft X-rays are produced from the inner-most regions of the hot accretion disk due to thermal radiation, while hard X-rays, up to a few 100 keV, could be produced by inverse Comptonisation of the disk photons by an optically thin hot electron cloud known as ``Corona'' (Done \emph{et al}. 2007). Radio and sub-mm emissions are believed to originate through Synchrotron processes in a bipolar Jet (Falcke \& Biermann 1996, 1999). Synchrotron emission has also been found to contribute in hard X-rays in some sources (Markoff \emph{et al}. 2001; Vadawale \emph{et al}. 2001). The origins of UV, optical, and IR emission are difficult to discern as there are multiple contenders for the same. The thin accretion disk, which is almost always approximated as a multi-temperature blackbody, emits X-rays close to the compact object (within tens of gravitational radii) owing to its inner temperature reaching as high as $\sim 10^7\, K$ and in longer wavelengths as one moves farther out (Shakura \& Sunyaev 1973). This emission is further escalated by irradiation of the outer parts of the disk by X-rays from the inner disk and/or back scattered photons from the Corona (Hameury 2020). The strength of this irradiation depends on the geometry of the Corona and scale height of the outer disk (Cunningham 1976; van Paradijs \& McClintock 1994). The companion star, which in case of low-mass X-ray binaries is a late type M, K, or G class star, peaks in optical or IR (OIR) wavelengths. This emission from the companion is also enhanced by irradiation of X-rays from the accretion disk by a few to several percentage. Synchrotron emission from relativistic jets can also dominate the OIR flux (Corbel \& Fender 2002; Russell \emph{et al}. 2006). In fact, the crucial break frequency dividing the optically thick and thin portion is believed to lie in the OIR bands which can help in quantifying the total energy content of a jet (Russell \emph{et al}. 2013). However, this break frequency is observed only in a very few sources with good significance (Coriat \emph{et al}. 2009). 

Deciphering the correct emission mechanism behind UV, optical, and IR emission remains challenging. Many techniques have been developed over the years to ascertain their true origin. Emissions from jet are known to show rapid variability in short time scales whereas those from disk or irradiated disk remain stable (Gandhi \emph{et al}. 2008). High cadence observations in OIR bands can shed some light on the mechanism involved in its emission (Curran \& Chaty 2013, Kosenkov \emph{et al}. 2020). The study of correlations between contemporaneous OIR and X-ray emission can also help in picking the dominant mechanism (Russell \emph{et al}. 2006; Bernardini \emph{et al}. 2016). Fitting a simultaneous broadband spectral energy distribution (SED) spanning radio to X-rays is also a well-established method. Furthermore, there are several models which attempt to explain the low frequency radiation by Synchrotron emission from a hot plasma above the disk (Veledina \emph{et al}. 2013). In these models, the hard X-ray emission is produced by thermal Comptonisation of soft Synchrotron photons. Optical excess is also thought to be produced from magnetic reconnections in flares on the disk (Merloni \emph{et al}. 2000) or from gravitational energy release near the circularisation radius (Campana \& Stella 2000). The problem, however, remains that of degeneracy wherein multiple models are able to satisfactorily explain the emissions whereas the available data are unable to discriminate between various competing models. Here we attempt to discern the origins of the near infrared (NIR), optical, and UV emissions observed during outburst of a black-hole binary GRS 1716--249 by evaluating the broadband SED.  

GRS 1716--249 (aka GRO J1719--24 and Nova Oph 1993; hereafter, to be referred as GRS1716) went into outburst on 18 December 2016 after more than 20 years of quiescence (Negoro \emph{et al}. 2016). During its discovery outburst in 1993, GRS1716 was measured to be located at a distance of $2.4 \pm 0.4\, kpc$ along with harboring a K type (or late) companion in a $14.7\, hr$ orbit (della Valle \emph{et al}. 1994). The lower limit on black hole mass was estimated to be $4.9\, M_\odot$ (Masetti \emph{et al}. 1996). During 2016, GRS1716 underwent a ``failed'' outburst wherein it didn't transition into high/soft or soft-intermediate state (Bassi \emph{et al}. 2019). It also lied in the outlier branch of radio/X-ray correlation plot with $L_R \propto L_X^{1.4}$ \footnote{$L_R$ and $L_X$ refer to radio and X-ray luminosity respectively.}. Bharali \emph{et al}. (2019) found minimal to no disk truncation along with detecting type C quasi periodic oscillations whose frequency increased with time. Jiang \emph{et al}. (2020) also concluded similarly about the inner disk radius along with providing constraints on the disk density. With joint \emph{Swift} and \emph{NuSTAR} spectroscopy, Tao \emph{et al}. (2019) constrained the spin and inclination of the source. The spin was constrained to a high value with $a \geq 0.92$ and inclination was estimated to lie within $40^\circ - 50^\circ$.  

\begin{figure}
\includegraphics[width=.95\columnwidth]{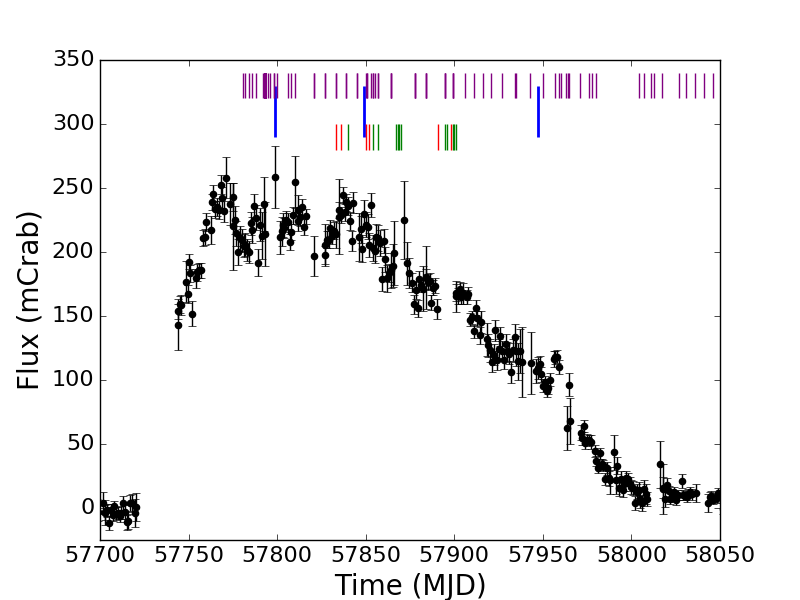}
\caption{MAXI on-demand lightcurve of GRS1716 in $2 - 20\, keV$ band (black circles; Matsuoka \emph{et al}. 2009). The blue vertical bars mark the three epochs of \emph{AstroSat} observations. UVOT observations in all six bands are depicted with violet bars. Optical (B,V,R,I) and NIR (J,H,Ks) observations from Mt. Abu are represented with red and green vertical bars.}
\label{maxilc}
\end{figure}

GRS1716 was also observed by \emph{AstroSat} as a Target of Opportunity on three epochs - 1) 15 Feb 2017 (57799), 2) 6 Apr 2017 (57849), and 3) 13 Jul 2017 (57947). It was also observed from Mount Abu Infrared Observatory during March to May 2017 for 24 nights in optical and NIR bands.  We present the joint spectral analyses of all three instruments of \emph{AstroSat} along with a multi-wavelength SED study to find the origin of NIR/optical/UV emission. The UV data were observed from \emph{Swift}/UVOT (Ultra-violet and optical Telescope). We have also utilised radio data from the Australian Large Baseline Array in the SED. The observation logs for this work are noted in Table \ref{tab:mirologs}. All the observations are further marked alongside the full MAXI lightcurve in Figure \ref{maxilc}.   

\section{Observations and Data Reduction}

\begin{table*}[htb]

\tabularfont
\caption{List of all observations used in this work. The wavelengths in UV, optical and NIR bands corresponds to the filters in the third column.}
\label{tab:mirologs}
\begin{tabular}{ccccc}

\topline
Bands & Observatory & Instrument & Energy/Wavelength & Date$^\star$ \\
\hline
X-ray & \emph{Astrosat} & SXT, LAXPC, CZTI & $1-120 \, keV$ & 15 Ferbruary, 06 April, 13 July \\
UV & \emph{Swift} & UVOT (W2,M2,W1) & 1928, 2246, 2600 \AA & 28 January - 13 August  \\
Optical & \emph{Swift} & UVOT (U,V,B) & 3465, 4392, 5468 \AA & 31 January - 20 October  \\
Optical & \emph{MIRO} & CCD (B,V,R,I) & 4353, 5477, 6349, 8797 \AA  & 22 March - 28 May  \\
NIR & \emph{MIRO} & NICS (J,H,Ks) & 1.25, 1.64, 2.15 $\mu m$ & 17 April - 25 May  \\
Radio$^\dagger$ & \emph{LBA} & - & $8.4\, GHz$ & 22 April \\



\hline
\end{tabular}
\tablenotes{$^\star$ All observations were made in the year 2017.}
\tablenotes{$^\dagger$ The data for the radio observation was reported by Bassi \emph{et al.} (2019)}
\end{table*}

\subsection{Soft X-ray Telescope (\emph{SXT}/AstroSat)}

SXT (Singh \emph{et al}. 2016, 2017) data were analysed with the standard analysis software and other auxiliary tools developed by the Payloads Operations Center (POC\footnote{www.tifr.res.in/$\sim$astrosat\_sxt}). The tool \texttt{sxtpipeline} was run to generate orbit-wise Level 2 event files. This extraction takes care of most of the elementary data cleaning and good-time interval (GTI) selection. The event files for every orbit were then merged using the script \texttt{sxtevtmergerjl}. Final products were generated using the FTool \texttt{xselect} after incorporating custom GTIs to remove flaring regions and drop outs from the light curve. SXT spectra are known to be affected by pile up for rates above $\sim 40\, counts/s$ (or 200 mCrab) in Photon Counting (PC) mode. The count rates for GRS1716 during the first two epochs were $\sim 42$ and $\sim 59\, counts/s$ respectively. In order to verify the presence of pile up, annular regions with outer radius fixed at $12'$ and inner radius varying from $1'$ upwards were used to generate spectra. Each of these spectra was fitted with an absorbed multi-color disk and powerlaw model and then the variation of photon index studied. The spectra were found to be piled up and thus annular regions with inner radii of $1'$ and $2.5'$ were selected for source extraction for epoch 1 and 2 observations respectively. The epoch 3 spectrum, with a count rate of $\sim 28 \, counts/s$, was not piled up and hence, a circular region of $15'$ radius was opted. To incorporate the changes in the effective area due to the annular region and vignetting caused by the off-axis positioning of the PSF (Point Spread Function), the default ARF (Auxiliary Response File) was scaled using the script \texttt{sxtARFModule}. The response matrix and the background spectra were provided by the POC. The spectra were grouped to have a minimum 30 counts per energy bin and a systematic error of 2\% was added.                

\subsection{Large Area X-ray Proportional Counter (\emph{LAXPC}/AstroSat)}

The analysis of LAXPC (Yadav \emph{et al}. 2016a, 2016b; Agrawal \emph{et al}. 2017) data were carried out using the Format A - \texttt{LAXPCsoftware\_Aug4} - package\footnote{http://astrosat-ssc.iucaa.in/}. LAXPC30 module was affected by gas leakage resulting in continuous gain instability (Antia \emph{et al}. 2017). Hence, only LAXPC10 and LAXPC20 were chosen for spectral analysis in this work. All the layers of both the LAXPCs were opted to maximise the signal as there were more than 15\% of counts in the bottom four layers. Level 2 event files were generated using the tool \texttt{laxpc\_make\_event} which was followed by the usage of \texttt{laxpc\_make\_spectra} and \texttt{laxpc\_make\_backspectra} for source and background spectral extractions. The channels in the spectra were grouped by a factor of 5 and a systematic error of 2\% was added. 

\subsection{Cadmium Zinc Telluride Imager (\emph{CZTI}/AstroSat)}

The reduction of CZTI (Vadawale \emph{et al}. 2016; Bhalerao \emph{et al}. 2017) level 1 files to level 2 and final products were carried out using the tool \texttt{cztpipeline} (Version 2.1). Spectrum from only Quadrant 0 was used for analysis as the other quadrants are affected by higher systematics. The spectra were grouped to have a minimum of 30 counts per energy bin and no systematic error was added.

\subsection{Ultraviolet and Optical Telescope (\emph{UVOT}/Swift)}

\begin{figure}
\includegraphics[width=.95\columnwidth]{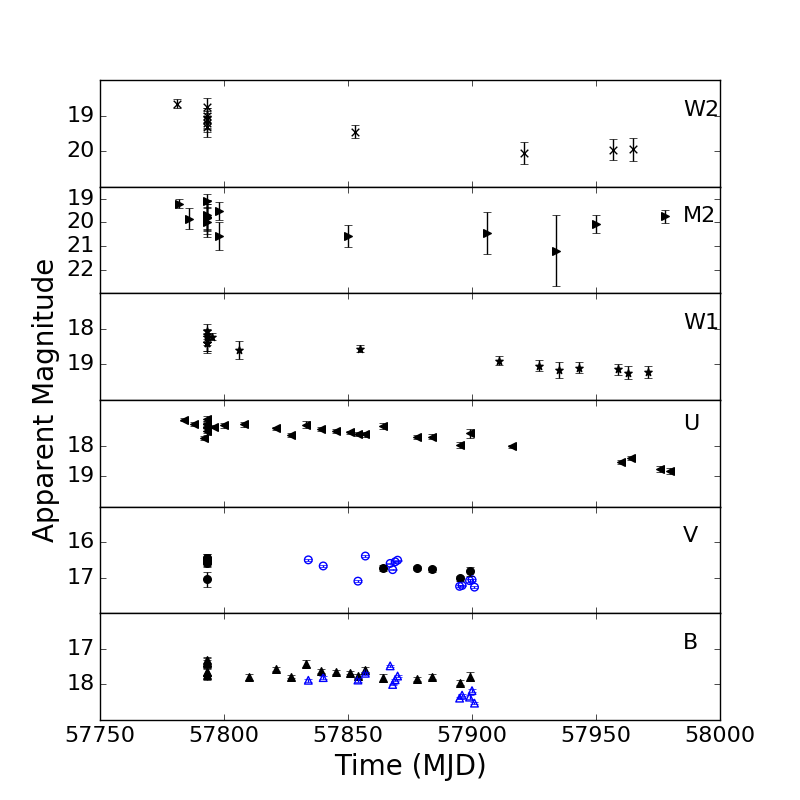}
\caption{Lightcurves in UV and optical bands using \emph{Swift}/UVOT. The top three panels display the lightcurves in UV filters (W2, M2, W1) while the bottom three show the optical lightcurves (U, V, B). The blue dots in V and B bands represent the same filters as observed from \emph{MIRO}.}\label{uvotlc}
\end{figure}

GRS1716 was observed several times during the outburst with UVOT onboard \emph{Swift} satellite (Breeveld \emph{et al}. 2011). Photometry for all the filters (W2, M2, W1, U, B, V) was done using the tool \texttt{uvotsource}. A circular region of $5''$ radius was considered for source while a source-free aperture of $20''$ radius was chosen for background extraction. The U, B, and V magnitudes of UVOT were converted to Johnson system using the conversion factors described by Poole \emph{et al}. (2008). The flux conversions were also done using the zero-point values for Vega flux given by Poole \emph{et al}. The lightcurves with all six UVOT filters are depicted in Figure \ref{uvotlc}.   

\begin{figure}
\includegraphics[width=.95\columnwidth]{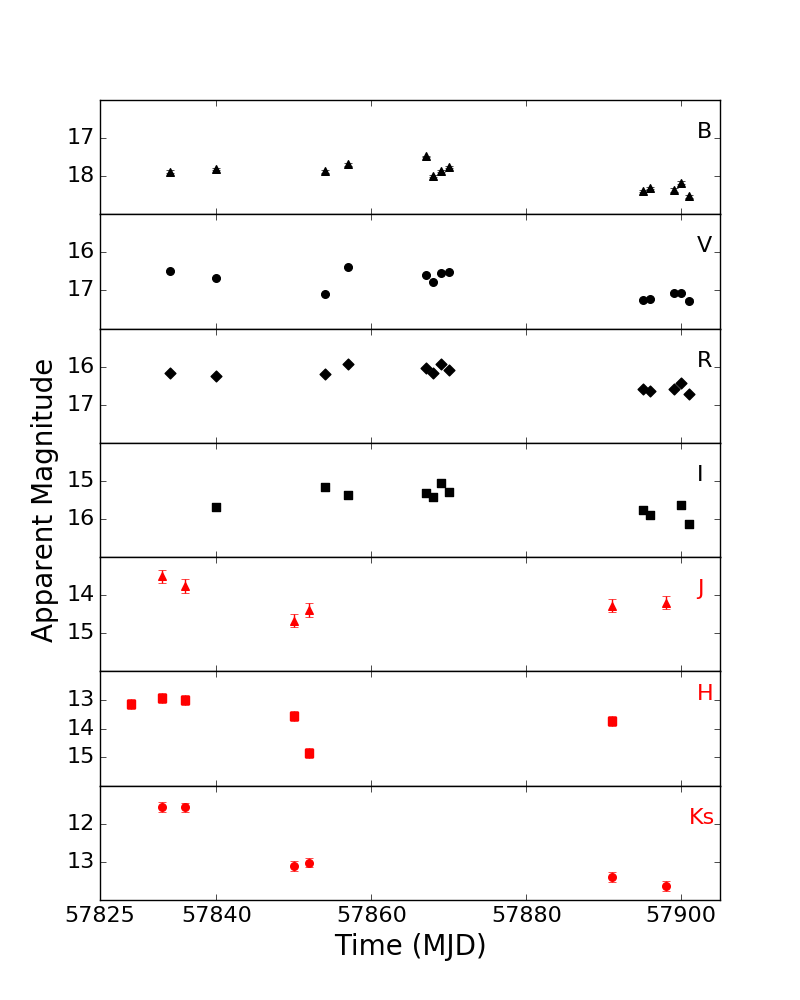}
\caption{Optical and NIR lightcurve of GRS1716 as observed from \emph{MIRO}. From top to bottom, the black points represent the optical filters of B, V, R, I respectively (also labelled on the plots). Similarly, the red points represent the NIR bands of J, H, and Ks respectively.}\label{mirolc}
\end{figure}

\subsection{Mount Abu Infrared Observatory (PRL)}

The Mount Abu Infrared Observatory (\emph{MIRO}), managed by Physical Research Laboratory, Ahmedabad, houses a 1.2 $m$ f/13 telescope with Cassegrain focus. Out of the many back end instruments in operation, two were used in this work - 1) Near Infrared Camera/Spectrograph (NICS; Anandarao \emph{et al}. 2008), and 2) Optical CCD. The optical filters (B,V,R,I) are in Johnson-Cousins photometric system and the NIR (J,H,Ks) filters are in MKO system. Four campaigns of observations were undertaken spanning 17 March to 28 May 2017 starting from a mid-plateau region to the first quarter of outburst decay of GRS1716 (Joshi \emph{et al}. 2017). 

Standard aperture photometry was carried out for optical and NIR observations using the \texttt{IRAF} package. NIR observations were acquired in 5 dithered positions separated by $\sim 30''$. These frames were median combined to produce a sky-frame which was subtracted from individual raw frames. Individual frames of the raw images for each day was subjected to bias subtraction for optical observations. Variable pixel response was corrected by the standard procedure of flat-field correction for both optical and NIR observations. The instrumental magnitudes of 3 to 4 field stars were then compared with corresponding apparent magnitudes from the standard catalogues viz. SDSS and 2MASS for optical and NIR, respectively, to find a zero point. The R and I band magnitudes of the comparison stars were in Sloan filters. The differences in Sloan and Johnson-Cousins filters were taken into account using transformation equations calculated by Jordi \emph{et al}. (2006). Similarly, the 2MASS magnitudes of the comparison stars were converted to MKO system\footnote{https://sites.astro.caltech.edu/~jmc/2mass/v3/transformations/}. This zero-point factor was, in turn, considered to calculate the apparent magnitude and standard deviation of the source. The Vega flux for magnitude to flux conversion for all 7 filters were taken from Bessell \emph{et al}. (1998). The optical and NIR lightcurves with MIRO are shown in Figure \ref{mirolc}.

\begin{figure}
\includegraphics[width=0.95\columnwidth]{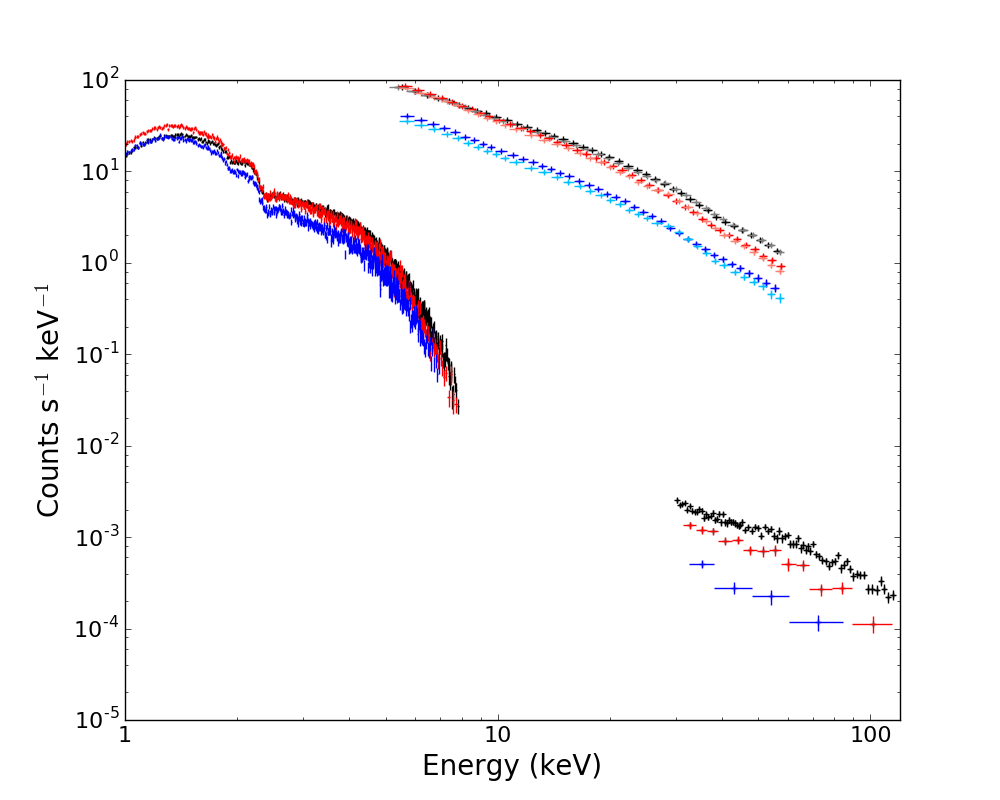}
\caption{\emph{AstroSat} spectra of all three epochs and instruments. From top to bottom - Epoch 1: 15 February 2017 (black), Epoch 2: 06 April 2017 (red), and Epoch 3: 13 July 2017 (blue). The spectrum LAXPC20 is represented by a lighter shade on all three days.}
\label{as1spec}
\end{figure}

\section{Analyses and Results}


\begin{figure*}
\centering
\includegraphics[scale=0.31]{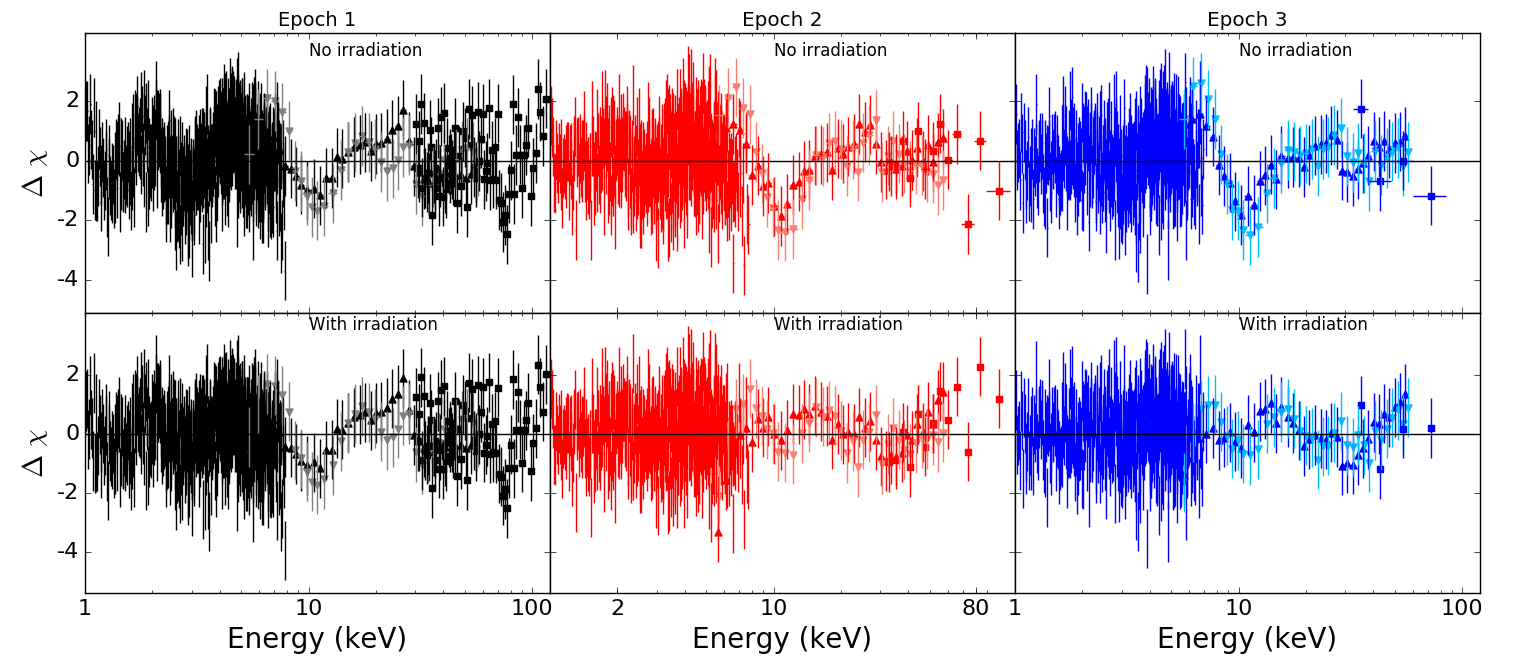}
\caption{The top panel displays the residuals of the \emph{AstroSat} spectra with \texttt{diskir} model assuming no irradiation in the inner disk for all three epochs. The bottom panel displays residuals after including irradiation. The color coding is the same as in Figure \ref{as1spec}. Dots, up-triangles, down-triangles, and squares are used to represent SXT, LAXPC10, LAXPC20, and CZTI respectively.}
\label{as1res}
\end{figure*}

\subsection{Joint AstroSat spectral fitting} \label{jointas1}

Spectral analysis was carried out by jointly fitting SXT, LAXPC10, LAXPC20, and CZTI using \texttt{xspec} (Arnaud 1996). The energy ranges of SXT and LAXPC spectra were restricted to $1 - 7\, keV$ and $5 - 60 \, keV$ respectively so as to avoid higher systematic errors outside these ranges. CZTI spectrum was fitted in the full range of $30 - 120 \, keV$. Thus, the combination of the three instruments resulted in a contiguous and wide energy coverage from 1 to 120 $keV$ (henceforth, the 4 \emph{AstroSat} spectra would be referred to as one X-ray spectrum). For SXT, an additional gain correction was added using the command \texttt{gain fit} in \texttt{xspec}. The best fit offset was found to be $\sim 40 \, eV$, for a unit slope, which improved the fits  significantly\footnote{https://www.tifr.res.in/$\sim$astrosat\_sxt/instrument.html}. To adjust for the cross-calibration discrepancy among the different instruments, a \texttt{constant} was multiplied to the model. It was fixed to unity for LAXPC10 and was left free to vary for the rest of the instruments. The average value of the best-fit \texttt{constant} factor for SXT and CZTI varied around $\approx 21\%$ of LAXPC10's factor whereas LAXPC20 varied between $\approx 5\%$. These are within the expected uncertainties in the effective areas of the three instruments.

The \emph{AstroSat} spectrum was fitted with an absorbed multi-temperature accretion disk (\texttt{diskbb}; Mitsuda \emph{et al}. 1984, Makishima \emph{et al}. 1986) and a thermal Comptonisation model (\texttt{nthComp}; Zdziarski \emph{et al}. 1996, {\.Z}ycki \emph{et al}. 1999) - \texttt{TBabs*(diskbb+nthComp)} - for all three epochs. The abundance for ISM absorption in \texttt{TBabs} was set to Wilms \emph{et al}. (2000). The fit for all epochs were statistically good with $\chi^2_\nu$ staying around 1.04. The hydrogen column density ($N_H$) was constrained to $\sim 0.7 \times 10^{22}\, cm^{-2}$, varying by 0.1 across the three epochs. The inner disk temperature and photon index were constrained to $\sim 0.3\, keV$ and $\sim 1.7$ respectively. The spectrum for all three epochs is depicted in Figure \ref{as1spec} and the residuals of the above fit are shown in the top panel of Figure \ref{as1res}. Despite good statistics, the residuals seem to be of somewhat wavy nature indicating the presence of reprocessed Coronal emission from the disk. Although reflection is a common candidate for reprocessing, we did not find telltale features like broad Fe line or Compton hump in the residuals. 

Nevertheless, we checked for the presence of reflection in the spectra by adding a relativistic reflection component - \texttt{relxillCp} - to the existing model (Dauser \emph{et al}. 2014, Garc\'ia \emph{et al}. 2014). The photon index and cutoff energy in \texttt{relxillCp} were tied to the corresponding parameters in \texttt{nthComp}. The emissivity index was fixed to the Newtonian value and the Fe abundance was set to that of the Sun. The spin parameter was fixed to 0.998 and the inner-disk radius was left free to vary. The possible values of inclination vary between $30^\circ - 50^\circ$ as reported by various authors (Bharali \emph{et al}. 2019, Tao \emph{et al}. 2019). Since inclination could not be constrained, we fixed it to a rough average of $40^\circ$. Rest of the parameters were left to float. There was only a marginal improvement in fit compared to the previous model with $\Delta \chi^2$ of 7.7, 10.4, and 24.1 per degree of freedom (dof) for epochs 1, 2, and 3 respectively. The inner radius could not be strongly constrained but hinted towards a possible disk truncation. Fixing spin to intermediate and Schwarschild values like $a=0.7,\, 0.3\, \& \,0$ also did not have any effect. The cutoff energy was only weakly constrained in the first epoch at $74.2^{+52.8}_{-14.3}$. For the other two cases, it pegged at the maximum limit and could not be constrained. This is in contrast to the values obtained by Bassi {et al}. (2019) who strongly constrained $kT_{e}$ to $\sim 48$ and $\sim 52\, keV$ for the first two epochs, and weakly at $\sim 74\, keV$ for the third. This is probably due to competition of the Compton hump to fit the curvature around $40-50\, keV$ which resulted in the cutoff energy being unconstrained. 

In the hard state of black-hole binaries, thermalisation of Comptonised photons in the inner disk can also substantially contribute to the disk emission (Gierli\'nski \emph{et al}. 2008). To test this, the spectrum was fitted with the \texttt{diskir} model (Gierli\'nski \emph{et al}. 2008). \texttt{diskir} is a hybrid of both blackbody and Comptonisation components. It parameterises irradiation by two additional components - 1) Fraction of Compton tail that is thermalised ($f_{in}$), and 2) Radius of Compton illuminated disk ($r_{irr}$); along with calculating the ratio of luminosities in Compton tail and unilluminated disk ($L_c/L_d$). It also has two more parameters for effects of irradiation at the outer disk. These parameters were frozen to nominal values as they would not affect X-ray emission and would be considered in the next exercise where broadband SED fitting is undertaken. First, irradiation was turned off by freezing $f_{in}$ to 0. The best-fit parameters and fit statistics were almost identical to the previous fits with \texttt{TBabs*(diskbb+nthComp)}. The values of $L_c/L_d$ were constrained around 5 indicating strong reprocessing which would also affect the thermal disk emission (Gierli\'nski \emph{et al}. 2008). To include this effect, $f_{in}$ was fixed to 0.1 and $r_{irr}$ was left free to vary while fitting. For epochs 2 and 3, the fits improved significantly while for epoch 1 it deteriorated a bit. $f_{in}$, for epoch 1, was found to be smaller by roughly a factor of 3. $r_{irr}$ was constrained to $\sim 1.01\, R_{in}$ (where $R_{in}$ is the inner-disk radius). The improvement in fit after including irradiation was verified by F-Test, wherein the probability that this advancement would be random was found to be less than $10^{-15}$ for all three epochs. The best-fit parameters are listed in Table \ref{tab:diskir} and the residuals are represented in bottom panel of Figure \ref{as1res}. There is a clear improvement in the fit after including irradiation. Fits with disk irradiation were also statistically better compared to reflection model with $\Delta \chi^2$ decreasing as 32.2, 11.6, and 26.1 per dof for the three epochs. Thus, the data favors the case for disk irradiation and does not require any reflection component. In reality, however, the spectrum could have contribution from both reflection and irradiation. However, deciphering the exact fraction of contribution from each of the two would be extremely difficult given the modest resolutions of SXT and LAXPC.

\begin{figure}
\includegraphics[width=.95\columnwidth]{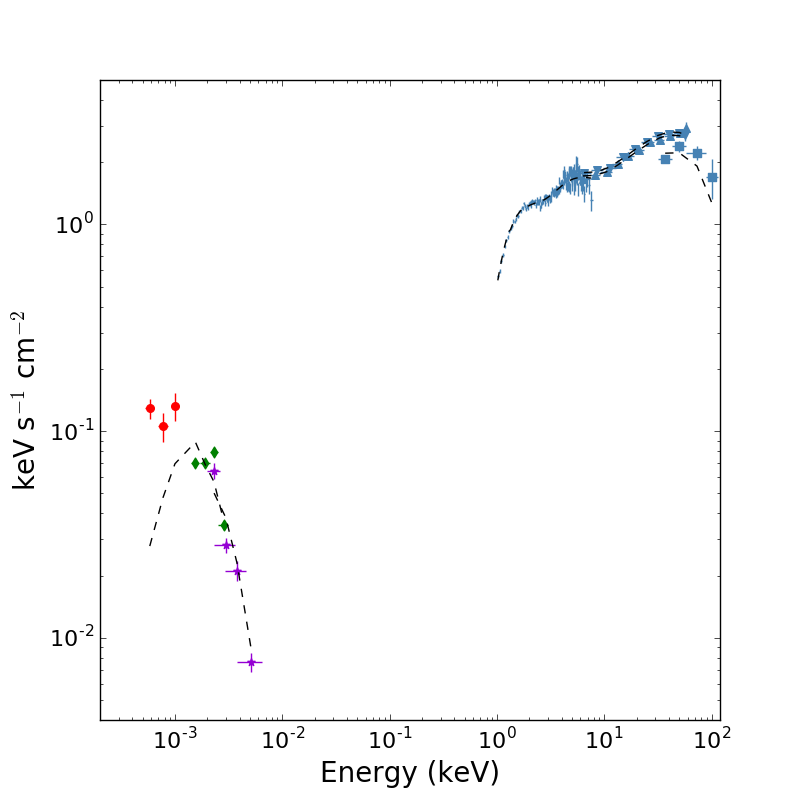}
\caption{SED of GRS1716 after fitting with \texttt{TBabs*diskir} to the quasi-simultaneous observations on epoch 2. The X-ray spectrum, in steelblue color, is unfolded and rebinned by a factor of 4 for clarity. The SXT, LAXPC10, LAXPC20, and CZTI are represented by dots, up-triangles, down-triangles, and squares respectively. In lower energies, the NIR spectrum is represented with red circles, optical points with green diamonds, and UV points with violet stars. The black dashed line represents the best-fit model.}
\label{excess}
\end{figure}

\subsection{Broadband Spectral Energy distribution}

Although there were many observations of GRS1716 in the low energy bands, there weren't any strictly simultaneous with \emph{AstroSat} observations. It was only during the second epoch (06 April) of \emph{AstroSat} when a number of observations with \emph{MIRO} and UVOT were in temporal proximity with it. The closest were the NIR observations which were made on 7 April while the optical measurements were made a few days later on 11 April. Similarly, the UV observations in W1, U, B, and V bands were scattered within a few days of 6 April. Moreover, around the date of the X-ray observation, the optical and UV flux did not vary significantly allowing multi-waveband spectroscopy (Figures \ref{mirolc} and \ref{uvotlc}). The other two UV bands of UVOT (M2 and W2) were much farther away in time from the X-ray observation and also suffered from heavy extinction leading to huge uncertainty in flux measurement. Therefore, they were not included in the SED.   

As hinted in section \ref{jointas1}, the model \texttt{diskir} also calculates the effect of irradiation in the outer disk where the emission is predominantly in UV, optical, and NIR bands (Gierli\'nski \emph{et al}. 2009). This is parameterised as $f_{out}$ and $R_{out}$ which represents the fraction of total flux thermalising the outer disk and the outer-disk radius respectively. The NIR, optical, and UV magnitudes were converted to flux in physical units and then incorporated into PHA files, one each for \emph{MIRO} and UVOT. Diagonal response matrices were created such that the convolved model would be in the same unit as the spectra. All spectra were loaded into \texttt{xspec} and fitted simultaneously with the model \texttt{TBabs*redden*diskir}. Since \emph{MIRO} and UVOT spectra were not corrected for interstellar absorption, the \texttt{xspec} routine \texttt{redden} was employed to calculate extinction (Cardelli \emph{et al}. 1989). It has one free parameter, the color excess ($E_{B-V}$), and it was left to float. $N_H$ and $E_{B-V}$ were fixed to 0 for UV/optical/NIR and X-ray spectrum, respectively, to avoid inter-mixing of the effects. $E_{B-V}$ was constrained to $0.75 \pm 0.04$ and $N_H$ was constrained to $0.70 \pm 0.04 \times\, 10^{22}\, cm^{-2}$ making the ratio $E_{E-B}/N_H$ $\sim 1.1 \times 10^{-22} \, cm^2\, mag$. This ratio is fully consistent with the recent findings of Lenz \emph{et al}. (2017). The cumulative line of sight galactic reddening
is $\sim 0.93$ (Schlafly \& Finkbeiner 2011) and our result is also consistent with this limit. The best-fit value of $f_{out}$ was $0.04 \pm 0.01$ while $R_{out}$ was constrained to $\sim 10^{5.5} R_{in}$. Although the irradiated disk model explained the optical and UV bands well, it failed to account for the excess in the NIR emission (see Figure \ref{excess}). To verify whether the NIR excess could be due to intrinsic absorption we multiplied another reddening component to the model. We fixed one $E_{B-V}$ to the best-fit value of 0.75 and let the other to float. The extra reddening component could not explain the NIR excess and was weakly constrained to a small value of $\sim 0.05$.
   

\begin{table*}[htb]
\tabularfont
\caption{Best-fit parameters of the joint X-ray fit with SXT, LAXPC, and CZTI using \texttt{TBabs*diskir} model.}\label{tab:diskir}
\begin{tabular}{cccccccccc}
\topline
Date & \textbf{$N_H$} &  \textbf{$kT_{in}$} & \textbf{$\Gamma$} & \textbf{$kT_e$} & \textbf{$L_c/L_d$} & \textbf{$f_{in}$} & \textbf{$r_{irr}$} & $norm$ & $\chi^2_{\nu}$ \\
(MJD) & ($\times 10^{22} cm^{-2}$) & ($keV$) & & ($keV$) & & & ($R_{in}$) & ($\times 10^3$) & ($\chi^2/dof$) \\
\midline
57799 & $0.64 \pm 0.05$ & $0.47 \pm 0.02$ & $1.60 \pm 0.01$ & $46^{+7}_{-5}$ & $8.2 \pm 0.6$ & $0.03^\star$ & $1.012 \pm 0.002$ & $2.26^{+0.79}_{-0.49}$ & 0.93 \\
57849 & $0.66 \pm 0.02$ & $0.44 \pm 0.01$ & $1.68 \pm 0.01$ & $22 \pm 1$ & $2.54 \pm 0.04$ & $0.1^\star$ & $1.011 \pm 0.001$ & $5.80^{+0.04}_{-0.03}$ & 0.92 \\
57947 & $0.52 \pm 0.02$ & $0.48 \pm 0.01$ & $1.65 \pm 0.01$ & $23^{+1}_{-2}$ & $2.73^{+0.05}_{-0.14}$ & $0.1^\star$ & $1.010 \pm 0.001$ & $1.87^{+0.03}_{-0.01}$ & $0.91$ \\
\hline
\end{tabular}
\tablenotes{$^\star$ Fixed during fit. }
\end{table*}  

\section{Discussion}

GRS1716 exhibited ``failed'' outburst and never transitioned to the canonical soft state (Bassi \emph{et al}). This was also suggested by the broadband rms variability of the source in the $3-30\, keV$ band. For epoch 1, the variability was $\approx 24\%$ whereas for epochs 2 \& 3 the variability remained around $20\%$. During the three \emph{AstroSat} observations, spanned across $\sim 5$ months, GRS1716 remained in a powerlaw dominant state with the luminosity ratio ($L_c/L_d$) remaining $> 2$. The spectrum of the source was significantly affected by irradiation of the back scattered Compton flux in the inner disk (Gierli\'nski \emph{et al}. 2008). For hard states of black-hole binaries, the fraction of the thermalising flux ($f_{in}$) is about 0.1 (Poutanen \emph{et al}. 1997). This fraction is a function of the geometry of the electron cloud and angle-averaged albedo of the thin disk. While epochs 2 and 3 confirmed the expected value of 0.1, $f_{in}$ for the first epoch was constrained to 0.03. This suggests a possible change in geometry, and hence covering fraction, of the overlying electron cloud. The complete outburst of GRS1716 is marked by 3 small softening episodes, also characterised by increase in flux. The epoch 2 \emph{AstroSat} observation (MJD 57849) was done 5 days before the second peak (MJD 57854). The effect of this softening was reflected in the spectral fits wherein, there was an increase in the disk flux and decrease in the luminosity ratio (Table \ref{tab:diskir}).

\begin{figure*}
\includegraphics[scale=0.4]{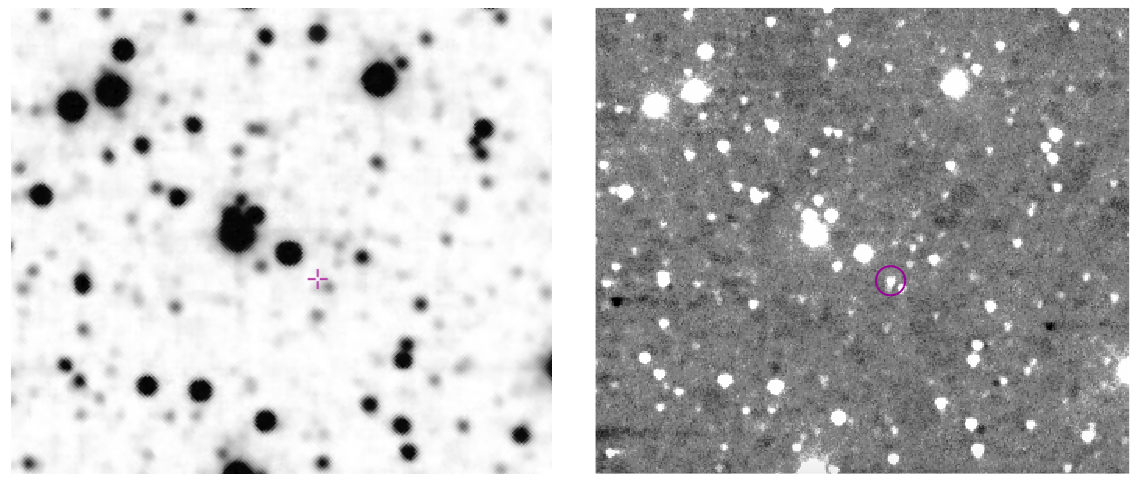}
\caption{Left panel shows the 2MASS image of the source during quiescence phase. The purple crossbar marks the source position. The right panel shows the H band image of the same field of view as observed with NICS instrument at \emph{MIRO} during the outburst. Inside the purple circe, a bright object is clearly seen.}
\label{nics2m}
\end{figure*}

We monitored GRS1716 in the optical and NIR bands from \emph{MIRO} during mid-March to May 2017. The optical lightcurve (Figure \ref{mirolc}) remained constant throughout the observations while the NIR lightcurve was marked by a drop in flux (1 - 2 magnitudes) around MJD 57850. The UVOT lightcurve (Figure \ref{uvotlc}), which spanned a longer duration, displayed constant flux in all bands up to MJD 57900 and a gradual decrease in the UV bands thereafter. Using quasi-simultaneous X-ray, UV, optical and NIR spectra we carried out a broadband spectral study to decipher the origin of the low energy emission using an irradiated disk model. The irradiated disk perfectly explained the optical and UV flux while slightly underestimating the NIR spectrum (Figure \ref{excess}). Without irradiation, the model underpredicted the flux by almost 3 orders of magnitude in all the bands. From the best-fit norm of \texttt{diskir} ($4181^{+1451}_{-887}$), the inner-disk radius was calculated to be $\sim 22\, km$, which was consistent with radii obtained by Bassi \emph{et al.}. The best-fit value of $f_{out}$ was found to be $0.03 \pm 0.01$ and that of $R_{out}$ was constrained to $\sim 6.27 \times 10^{6}\, km$. This $R_{out}$ is slightly larger than the Roche lobe radius of $\sim 1.56 \times 10^{6}\, km$  estimated from the reported orbital parameters and assuming a $10\, M_\odot$ black hole and $45^\circ$ disk inclination\footnote{The Roche lobe radius - distance of the inner Lagrangian point to the primary object (black hole) - was estimated using the following tool: http://www.orbitsimulator.com/formulas/LagrangePointFinder.html}. Considering the uncertainties involved in the above calculation, we can infer that the outer disk has to be as big as the Roche lobe in order to describe the UV/optical spectrum. Trying to increase $R_{out}$ to account for the NIR spectrum would make the disk unrealistically large and also overestimate the optical spectrum. 

A natural alternative for the this excess is emission from a secondary star. We searched for the images of the source in all-sky surveys to quantify the flux during quiescence which would be predominantly from the companion. However, the 2MASS H-band image of the field does not have any object at the position of GRS1716 (Figure \ref{nics2m}). The H-band image from \emph{MIRO}, on the other hand, shows a bright object at the source position. Hence, the NIR brightening which the source had undergone during the outburst is not due to the secondary star. It is also possible for the emissions from the binary to be absorbed and re-emitted in the IR regime from a dust envelope/cloud covering the binary. Taranova \& Shenavrin (2001) reported such a scenario where the X-ray binary XTE J1118$+$480 showed excess in mid-IR regime which could be explained by a 900 $K$ circumstellar dust envelope. To test this hypothesis, we added the \texttt{bbodyrad} model to the SED only in the UV/optical/NIR region. We fixed the temperature to $900\, K$ and fitted for the norm. The best-fit radius of the cloud was found to be $\sim 3.3 \times 10^8 \, km$. The cloud, if present, could as well be hotter and smaller in size or cooler and larger. Without observations in longer wavelengths, it is difficult to constraint any of these properties robustly. Although the primary source of heating of a bright dust cloud is disk emission during an outburst, it can also be moderately excited by emission from the secondary star. The object was, however, not detected in the mid-IR bands of the WISE (Wide-field Infrared Survey Explorer) catalogue observed during the 2010-11 epoch. Moreover, the source was also not detected in NIR, or even optical bands, after we resumed observations post Monsoon during Sep-Oct 2017. During these months the source had decayed substantially in X-rays but was still bright in UV (see UVOT lightcurves in Figure \ref{uvotlc}) This suggests that the NIR and optical brightening was exclusively tied to the X-ray activity. The possibility of a NIR emitting dust cloud that engulfs the entire system is, therefore, highly speculative.

The other most viable candidate for the NIR excess is Synchrotron emission from a compact jet. As reported by Bassi \emph{et al}., GRS1716 was detected in radio wavebands throughout the outburst by ATCA, LBA, and VLA from 9 February 2017 till 13 August 2017 (see their Table 3). The closest radio observation to our SED was made on 22 April using LBA. The flux in the 8.4 $GHz$ band, with a bandwidth of 64 $MHz$, was reported to be $1.13 \pm 0.11\, mJy$. The observation before 22 April was made about two months earlier, on 9 February, during which the flux was $1.28 \pm 0.15\, mJy$. This means the flux would not have changed much during the intervening period. The correlation between radio and X-ray luminosities provide an useful tool to study the emission properties of black-hole binaries. These are known to follow two distinct powerlaw relations in the log-log space. The radio-loud systems follow a relation $L_R \propto L_X^{1.4}$ whereas the radio-quiet systems (the so-called ``outliers") follow $L_R \propto L_x^{0.6}$ (Corbel \emph{et al}. 2013). Using the radio luminosity from LBA in the 8.4 $GHz$ band ($L_R \approx 6.5 \times 10^{28}\, erg\, s^{-1}$) and the X-ray luminosity in the $1-10\, keV$ band from \emph{AstroSat} ($L_X \approx 4.5 \times 10^{36}\, erg\, s^{-1}$) we obtained a radio/X-ray luminosity relation of $L_R \propto L_X^{1.45}$. Here, we have assumed a distance of $2.4\, kpc$ and a proportionality constant of 1.85 (Corbel \emph{et al}. 2013). Thus, GRS1716 adds to the pool of sources in the ``outlier" branch of the radio/X-ray plane, consistent with the findings of Bassi \emph{et al}.   

\begin{figure}
\includegraphics[width=.95\columnwidth]{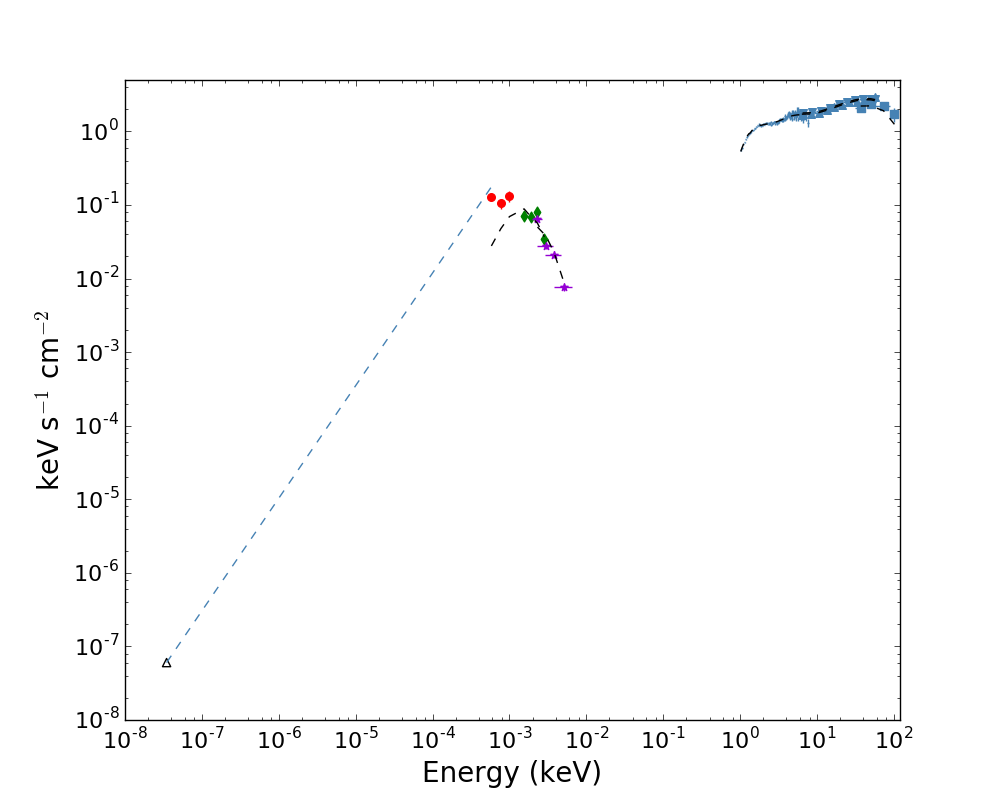}
\caption{Complete SED of GRS1716 along with the 22 April radio observation with LBA. The X-ray spectrum is rebinned by a factor of 4 for clarity. The X-ray, UV, optical, and NIR spectra are the same as in Figure \ref{excess}. The radio observation is marked with a black open triangle. The grey dotted line joining the radio and Ks band is just for representation and not a fitted model.}
\label{sed}
\end{figure}

To have a cursory idea of the full radio to X-ray SED, we also add the 22 April radio observation in the SED (Figure \ref{sed}). Unfortunately, with the data available with us, it is not possible to identify the exact position of the break frequency. We tried fitting the radio to NIR spectrum with a broken powerlaw but could not constrain the parameters, especially the break frequency. An approximate spectral index of the radio spectrum is $+0.5$, obtained by fixing the spectral break at the Ks band of the NIR spectrum. Such highly inverted optically thick part of the radio spectrum has been seen earlier for a few sources such as MAXI J1836--194 (Russell \emph{et al}. 2014), XTE J1118$+$480 (Fender \emph{et al}. 2001), etc. Although standard jet models, as that of Blandford \& K\"onigl (1979), assuming a conical geometry predict a shallower slope, steeper spectrum can be expected for a rapidly flaring jet geometry (Din{\c c}er \emph{et al}. 2018).          


\section{Conclusion}

We have presented the results of a multi-wavelength spectral analysis of the galactic X-ray binary GRS 1716--249 using data in X-rays from \emph{AstroSat}, NIR/optical from \emph{MIRO} and UV from \emph{Swift}/UVOT. Broadband X-ray spectral analysis of all three epochs of \emph{AstroSat} spectra show that the source was in a powerlaw dominant state. Irradiation of X-rays in the inner regions of the accretion disk significantly contribute to the soft X-ray flux of the source on all three epochs. Using multi-wavelength SED analysis, we found the optical and UV flux to originate from the irradiated outer accretion disk while parts of the NIR emission is most likely emitted from a jet.


\section*{Acknowledgements}

This work was supported by Physical Research Laboratory, a unit of Department of Space, Government of India. It uses data from the \emph{AstroSat} mission of the Indian Space Research Organisation (ISRO), archived at the Indian Space Science Data Centre (ISSDC). We thank the POCs of SXT (TIFR), LAXPC(TIFR), and CZTI (IUCAA) for verifying and releasing the data via ISSDC data archive and providing the necessary data analysis software through
the \emph{AstroSat} Science Support Cell. UVOT data were obtained from the HEASARC data archives and were analyzed using the information provided by the UK Swift Science Data Center (University of Leicester). This research has also made use of MAXI data provided by RIKEN, JAXA, and the MAXI team. SKR thanks Sushree S. Nayak for valuable feedback on the manuscript.




\begin{thebibliography}{100}



\bibitem{agrawal17}
Agrawal P. C. \emph{et al}. 2017, JApA, 38, 30

\bibitem{anandarao08}
Anandarao B. \emph{et al}. 2008, SPIE, 7014, 70142Y

\bibitem{antia17}
Antia H. M. \emph{et al}. 2017, ApJS, 231, 10

\bibitem{arnaud96}
Arnaud K. A. 1996, ASPC, 101, 17

\bibitem{bassi19} 
Bassi T. \emph{et al}. 2019, MNRAS, 482, 1587

\bibitem{bernardini16} 
Bernardini F. \emph{et al}. 2016, ApJ, 826, 149

\bibitem{bessell98}
Bessell M. S., Castelli F., Plez B. 1998, A\&A, 333, 231

\bibitem{bhalerao17}
Bhalerao V. \emph{et al}. 2017, JApA, 38, 31

\bibitem{bharali19} 
Bharali P. \emph{et al}. 2019, MNRAS, 487, 3150

\bibitem{blandford79}
Blandford R. D., K\"onigl A. 1979, ApJ, 232, 34

\bibitem{breeveld11}
Breeveld A. A. \emph{et al}. 2011, AIPC, 1358, 373

\bibitem{campana00} 
Campana S., Stella L. 2000, ApJ, 541, 849

\bibitem{cardelli89}
Cardelli J. A., Clayton G. C., Mathis J. S. 1989, ApJ, 345, 245

\bibitem{charles06} 
Charles P. A., Coe M. J. 2006, csxs.book, 39, 215

\bibitem{corbel02} 
Corbel S., Fender R. P. 2002, ApJL, 573, L35

\bibitem{corbel13}
Corbel S. \emph{et al}. 2013, MNRAS, 428, 2500

\bibitem{coriat09} 
Coriat M. \emph{et al}. 2009, MNRAS, 400, 123

\bibitem{cunnigham76} 
Cunningham C. 1976, ApJ, 208, 534

\bibitem{curran13} 
Curran P. A., Chaty S. 2013, A\&A, 557, A45

\bibitem{dauser14}
Dauser T. \emph{et al}. 2014, MNRAS, 444, L100

\bibitem{dellavalle94} 
della Valle M., Mirabel I. F., Rodriguez L. F. 1994, A\&A, 290, 803

\bibitem{dincer18}
Din{\c c}er T. \emph{et al}. 2018, ApJ, 852, 4

\bibitem{done07} 
Done C., Gierliński M., Kubota A. 2007, A\&ARv, 15, 1

\bibitem{falcke96} 
Falcke H., Biermann P. L. 1996, A\&A, 308, 321

\bibitem{falcke99} 
Falcke H., Biermann P. L. 1999, A\&A, 342, 49

\bibitem{fender01}
Fender R. P. \emph{et al}. 2001, MNRAS, 322, L23

\bibitem{gandhi08} 
Gandhi P. \emph{et al}. 2008, MNRAS, 390, L29


\bibitem{garcia14}
Garc\'ia J. \emph{et al}. 2014, ApJ, 782, 76

\bibitem{gierlinski08}
Gierli\'nski M., Done C., Page K. 2008, MNRAS, 388, 753

\bibitem{gierlinski09}
Gierli\'nski M., Done C., Page K. 2009, MNRAS, 392, 1106

\bibitem{hameury20}
Hameury J. M. 2020, AdSpR, 66, 1004

\bibitem{jiang20} 
Jiang J. \emph{et al}. 2020, MNRAS, 492, 1947

\bibitem{jordi06}
Jordi K., Grebel E. K., Ammon K. 2006, A\&A, 460, 339

\bibitem{joshi17} 
Joshi V., Vadwale S., Ganesh S. 2017, ATel, 10196, 1

\bibitem{kosenkov20}
Kosenkov I. A. \emph{et al}. 2020, A\&A, 638, A127

\bibitem{lenz17}
Lenz D., Hensley B. S., Dor\'e O. 2017, ApJ, 846, 38

\bibitem{makishima86} 
Makishima K. \emph{et al}. 1986, ApJ, 308, 635

\bibitem{markoff01} 
Markoff S., Falcke H., Fender R. 2001, A\&A, 372, L25

\bibitem{masetti96} 
Masetti N. \emph{et al}. 1996, A\&A, 314, 123

\bibitem{matsuoka09} 
Matsuoka M. \emph{et al}. 2009, PASJ, 61, 999

\bibitem{merloni00} 
Merloni A., Di Matteo T., Fabian A. C. 2000, MNRAS, 318, L15

\bibitem{mitsuda84} 
Mitsuda K. \emph{et al}. 1984, PASJ, 36, 741

\bibitem{negoro16} 
Negoro H. \emph{et al}. 2016, ATel, 9876, 1

\bibitem{poole08}
Poole T. S. \emph{et al}. 2008, MNRAS, 383, 627

\bibitem{poutanen97} 
Poutanen J., Krolik J. H., Ryde F. 1997, MNRAS, 292, L21

\bibitem{russell06} 
Russell D. M. \emph{et al}. 2006, MNRAS, 371, 1334

\bibitem{russell13}
Russell D. M. \emph{et al}. 2013, MNRAS, 429, 815

\bibitem{russell14}
Russell T. D. \emph{et al}. 2014, MNRAS, 439, 1390

\bibitem{schlafly11} 
Schlafly E. F., Finkbeiner D. P. 2011, ApJ, 737, 103


\bibitem{singh16}
Singh K. P. \emph{et al}. 2016, SPIE, 9905, 99051E 

\bibitem{singh17}
Singh K. P. \emph{et al}. 2017, JApA, 38, 29 

\bibitem{shakura73} 
Shakura N. I., Sunyaev R. A. 1973, A\&A, 500, 33

\bibitem{tao19} 
Tao L. \emph{et al}. 2019, ApJ, 887, 184

\bibitem{taranova01}
Taranova O. G., Shenavrin V. I. 2001, AstL, 27, 25

\bibitem{vadawale01} 
Vadawale S. V., Rao A. R., Chakrabarti S. K. 2001, A\&A, 372, 793

\bibitem{vadawale16}
Vadawale S. V. \emph{et al}. 2016, SPIE, 9905, 99051G

\bibitem{vanparadijs94} 
van Paradijs J., McClintock J. E. 1994, A\&A, 290, 133

\bibitem{vanparadijs95} 
van Paradijs J., McClintock J. E. 1995, xrbi.nasa, 58

\bibitem{veledina13} 
Veledina A., Poutanen J., Vurm I. 2013, MNRAS, 430, 3196

\bibitem{wilms00} 
Wilms J., Allen A., McCray R. 2000, ApJ, 542, 914

\bibitem{yadav16a}
Yadav J. S. \emph{et al}. 2016, SPIE, 9905, 99051D

\bibitem{yadav16b}
Yadav J. S. \emph{et al}. 2016, ApJ, 833, 27

\bibitem{zdziarski96} 
Zdziarski A. A., Johnson W. N., Magdziarz P. 1996, MNRAS, 283, 193

\bibitem{zycki99} 
{\.Z}ycki P. T., Done C., Smith D. A. 1999, MNRAS, 309, 561





\end{thebibliography}
\end{document}